\documentclass[review,3p,times,12pt]{elsarticle}
\usepackage{graphicx}
\usepackage{subcaption}
\usepackage{hyperref}
\usepackage{amsmath,amssymb,amsthm}
\usepackage{bigints}
\usepackage{amsfonts}
\usepackage{graphicx,epsfig, subcaption}
\usepackage{natbib}
\usepackage{enumitem}
\usepackage{indentfirst}
\usepackage{epstopdf}
\usepackage[font=small,labelfont=bf,labelsep=space]{caption}

\hypersetup{
    colorlinks = true,
    linkcolor = black,
    urlcolor  = black,
    citecolor = black,
    anchorcolor = black
}

\newtheorem{theorem}{Theorem}
\input epsf

\newcounter{saveeqn}

\begin{document}




\begin{frontmatter}
\title{\bf Conservation laws, a new class of group invariant solutions and its applications for Whitham–Broer–Kaup model}

\author[add1]{Sougata Mandal\corref{cor1}}
\ead{mandal.5@iitj.ac.in}
\author[add1]{Sukhendu Ghosh}
\cortext[cor1]{Corresponding Author}
\address[add1]{Department of Mathematics, Indian Institute of Technology Jodhpur, Rajasthan -342030, India}

\begin{abstract}
 The Whitham–Broer–Kaup (WBK) equations provide a fundamental framework for modeling shallow water wave dynamics, effectively capturing both nonlinear and dispersive effects. In this study, we construct a new class of analytical and numerical solutions for the WBK system using Lie symmetry analysis. By determining an optimal system of one-dimensional subalgebras, we obtain symmetry reductions that lead to new kinds of exact wave solutions expressed in hyperbolic, trigonometric, and rational forms. The influence of key physical parameters on wave structure is systematically explored, revealing their role in shaping the velocity and surface profiles of the waves. An important aspect of this work is the application of the WBK model to tsunami wave propagation, demonstrating its capability to simulate the generation, evolution, and spatial spreading of long surface waves in coastal regions. This highlights the practical relevance of the WBK equations in geophysical and oceanographic contexts. Additionally, employing the direct multiplier method, we derive a complete set of local conservation laws for the governing WBK model, ensuring the preservation of key physical properties. These findings enhance the understanding of shallow water wave behavior, unify existing research, and provide a framework for further exploration.
\end{abstract}

\end{frontmatter}

\section{Introduction}
Nonlinear shallow water wave equations play a crucial role in modeling fluid motion across various disciplines, including coastal engineering, oceanography, and hydrology \cite{ref1, ref3}. These equations are instrumental in understanding and forecasting natural events such as tsunami propagation, tidal movements, and riverine flows \cite{ref2, ref4}. These equations support engineers in designing flood control strategies, assessing dam-break scenarios, and simulating sediment transport. Their ability to capture essential fluid dynamics in a simplified form makes them a powerful tool for studying diverse geophysical processes.

The general mathematical representation of long-wave dynamics in shallow water is as follows:
\begin{align}\label{eq1}
u_t + u u_x + v_x + \alpha u_{xx} &= 0, \\ \nonumber
v_t + u_x v + u v_x - \alpha v_{xx} &= 0.
\end{align}
This set of equations models the movement of narrow water waves. In these equations, $x$ and $t$ denote the normalized space and time variables, $u(x,t)$ is the horizontal velocity field, and $v(x,t)$ describes the deviation of the water surface elevation from its undisturbed (equilibrium) level above a flat bottom, subscripts represent derivatives. The constant $\alpha$ is the diffusion parameter with varying effects on the system. The traveling wave solutions of Eq. \eqref{eq1} have been studied using the polynomial complete discriminant system method \cite{ref1}.

The Boussinesq equations form another important fundamental model for describing shallow water wave dynamics, particularly in scenarios involving weak nonlinearity and dispersion. Mathematically, this system is given by:
\begin{align}\label{eq2} 
& u_t + u u_x + v_x = 0, \\ & \nonumber v_t + (u v)_x + u_{xxx} = 0. \end{align}
These equations capture the interplay between nonlinearity and dispersion in wave propagation, making them highly relevant in various fields such as fluid dynamics, coastal engineering, and nonlinear optics. Exact solutions of Eq. \eqref{eq2} are derived using the $(G'/G)$-expansion method \cite{ref6}, while multisoliton solutions are obtained through the homogeneous equilibrium method \cite{ref7}.

The Whitham-Broer-Kaup (WBK) equations \cite{ref1} represent a more comprehensive model that includes both nonlinear effects and higher-order dispersion, and it is expressed as follows: 
\begin{align}\label{eq3}
    & u_t + u u_x + v_x + \alpha u_{xx} = 0, \\  & \nonumber
    v_t + (u v)_x + \tau u_{xxx} - \alpha v_{xx} = 0.
\end{align}
Here, $\tau$ is a constant that represents a different diffusion parameter.
This set of equations represents an important class of nonlinear evolution equations describing dispersion in shallow water waves. The WBK system effectively captures the balance between nonlinearity and dispersion, making it crucial for modeling wave interactions and soliton behavior. Beyond fluid mechanics, the WBK system finds applications in nonlinear optics, plasma physics, biological transport systems, and even traffic flow models. Its integrable structure enables soliton solutions, which are essential for understanding wave stability and long-term behavior in various physical systems. Several simplified forms of these equations emerge under specific parameter choices, and these have been extensively studied by numerous researchers. The following represent the different existing studies of the general model \eqref{eq3} for certain values of the flow parameters $\alpha$ and $\tau$. 
\begin{itemize}
    \item When  $\tau = 0$, the system \eqref{eq3} reduces to a long-wave system describing shallow water waves with diffusion \cite{ref1}, and the corresponding dynamics is governed by \eqref{eq1}.
    \item When $\alpha = 0$, $\tau = 0$, system \eqref{eq3} reduces to a dispersionless long-wave system modeling long waves in shallow water \cite{ref8}.
    \item When $\tau > 0$, system \eqref{eq3} reduces to the Kaup-Boussinesq shallow-water system \cite{ref9}.
    \item When $\tau = -\frac{1}{4}$, system \eqref{eq3} reduces to the Kaup-Boussinesq system \cite{ref10}.
    \item When $\tau = 1$, system \eqref{eq3} reduces to a variant Boussinesq system \cite{ref11}.
\end{itemize}
Thus, the equation \eqref{eq3} represents a generalized framework that encompasses several classical wave models as particular cases. Hence, deriving exact soliton solutions for this system is crucial for advancing the understanding of the dynamics of shallow water waves. In earlier studies, Changjing et al. employed the Darboux transformation method and the Wronskian determinant technique to investigate soliton solutions and their interaction properties for Eq. \eqref{eq3} \cite{ref12}. Zheng Zhi obtained exact solutions for Eq. \eqref{eq3} using the Tanh-function and Exp-function methods \cite{ref13}. Gao et al. \cite{ref27} examined the hetero-Backlund transformations and similarity reductions for system \eqref{eq3}. Oloniiju \cite{ref28} explored soliton solutions for the same system using the sine-Gordon expansion method. Similarly, Luo \cite{ref29} conducted a bifurcation analysis, offering insights into the system's intricate dynamics. Over the years, mathematicians have dedicated considerable effort to deriving exact rational and closed-form solutions for various nonlinear evolution equations. To address these challenges, several innovative analytical techniques have been developed, including the Backlund transformation \cite{ref17}, Darboux transformation \cite{ref18}, inverse scattering method \cite{ref19}, Hirota’s bilinear method, the modified generalized exponential rational function method \cite{ref20}, the unified technique \cite{ref22}, the singular manifold method \cite{ref23}, the Painlevé test \cite{ref24}, the Lie symmetry approach \cite{ref25}, and the generalized exponential differential function (GEDF) approach \cite{ref26}, etc.

Among these approaches, Lie symmetry analysis and optimal classifications have emerged as rigorous and effective methods for identifying various types of soliton solutions of complex nonlinear differential equations \cite{ref30, ref31}. Since the symmetry analysis of the considered governing equations \eqref{eq3} has been relatively unexplored, we are motivated by the aforementioned literature to employ Lie symmetry analysis on this model. In this study, by constructing a one-dimensional optimal system, we derive a new class of soliton wave solutions that combine hyperbolic, trigonometric, and rational functions for the WBK equations.
Additionally, we derive a set of local conservation laws. These new solutions and a set of conservation laws may provide deeper insights into shallow water wave behavior. The structure of the paper is as follows: Section \ref{sec1} presents the derivation of the symmetry transformations that leave the WBK equations invariant. An overview of the extended simplest equation method is provided in Section \ref{sec2}. In Section \ref{sec3}, we explore symmetry reductions and construct a new class of exact solutions for the WBK system. Section \ref{sec4} is devoted to the derivation of conservation laws using the direct multiplier method. Section \ref{sec55} discusses the application of the WBK system. Finally, Section \ref{sec5} concludes the study with a summary of the key findings.

\section{Symmetry Transformations} \label{sec1}
Let's consider a one-parameter group of continuous transformations acting on the variables \( x, t, u, v \), which leaves the WBK equations invariant, is generated by the following vector field:
\begin{equation}\label{eq4}
V = \xi_1(x, t, u, v) \frac{\partial}{\partial x} + \xi_2(x, t, u, v) \frac{\partial}{\partial t} + \eta_1(x, t, u, v) \frac{\partial}{\partial u} + \eta_2(x, t, u, v) \frac{\partial}{\partial v}.
\end{equation}
The associated third-order prolongation of this vector field is defined by:
\begin{equation}\label{eq5}
\text{pr}^{(3)} V = V + \eta_1^x \frac{\partial}{\partial u_x} + \eta_1^t \frac{\partial}{\partial u_t} + \eta_1^{xx} \frac{\partial}{\partial u_{xx}} + \eta_1^{xxx} \frac{\partial}{\partial u_{xxx}}
\eta_2^x \frac{\partial}{\partial v_x} + \eta_2^t \frac{\partial}{\partial v_t} + \eta_2^{xx} \frac{\partial}{\partial v_{xx}} + \eta_2^{xxx} \frac{\partial}{\partial v_{xxx}}.
\end{equation}
The prolonged coefficients can be computed using the following recurrence relations:
\begin{align*}
&\eta_k^S=D_S \eta_k- \sum_{j=1}^2U_{jx} D_S\xi_1-\sum_{j=1}^2U_{jt} D_S\xi_2,~~ \eta_k^{JS}=D_S\eta_k^J-\sum_{j=1}^2U_{jJx} D_S\xi_1-\sum_{j=1}^2U_{jJt} D_S\xi_2,
\end{align*}
where, $S$, $J$ stand for $x, t, xx, xxx$, and $k=1, 2$. Also, noted that $U_1 = u(x, t) $, $U_2 = v(x, t) $, and $D_S$ represents the total derivative operators. Moreover, the symmetry conditions for the system of PDEs are obtained by enforcing:
\begin{equation}\label{eq8}
\text{pr}^{(3)} V (\Delta_1)\big|_{\substack{\Delta_1 = 0, \ \Delta_2 = 0}} = 0, \quad \text{and} \quad \text{pr}^{(3)} V (\Delta_2)\big|_{\substack{\Delta_1 = 0, \ \Delta_2 = 0}} = 0,
\end{equation}
where, 
\begin{equation*}\label{eq6}
\Delta_1 = u_t + u u_x + v_x + \alpha u_{xx},~~
\Delta_2 = v_t + (u v)_x + \tau u_{xxx} - \alpha v_{xx}.
\end{equation*}
These symmetry conditions \eqref{eq8} give the following set of partial differential equations, which are used to determine the infinitesimals $\xi_j, \eta_j$ for $j=1, 2$ as follows:
\begin{align}\label{eq9}
&  \nonumber \frac{\partial \xi_2}{\partial x} = 0,~ \frac{\partial \xi_1}{\partial x} = -\frac{1}{2} \frac{\eta_2}{v},~
\frac{\partial \eta_1}{\partial x} = 0,~ \frac{\partial \eta_2}{\partial x} = 0, \frac{\partial \xi_2}{\partial t} = -\frac{\eta_2}{v},~ 
\frac{\partial \xi_1}{\partial t} = \frac{1}{2} \frac{2\eta_1 v - u\eta_2}{v},~ 
\frac{\partial \eta_1}{\partial t} = 0, \\
&~ \frac{\partial \eta_2}{\partial t} = 0,  \frac{\partial \xi_2}{\partial u} = 0,~ \frac{\partial \xi_1}{\partial u} = 0,~
\frac{\partial \eta_1}{\partial u} = \frac{1}{2} \frac{\eta_2}{v},~
\frac{\partial \eta_2}{\partial u} = 0,  \frac{\partial \xi_2}{\partial v} = 0,~ \frac{\partial \xi_1}{\partial v} = 0,~
\frac{\partial \eta_1}{\partial v} = 0,~ \frac{\partial \eta_2}{\partial v} = \frac{\eta_2}{v}.
\end{align}

Solving the above set of partial differential equations \eqref{eq9}, one can obtain the following solutions:
\begin{align*}
\eta_1 = \frac{1}{2}C_1 u + C_3,~ \eta_2 = C_1 v,~
 \xi_1 = C_3 t - \frac{1}{2}C_1 x + C_4,~\xi_2 = -C_1 t + C_2.
\end{align*}
where, $C_i$ for $i=1(1)4$ are arbitrary constants. This set of solutions gives the following theorem.

\begin{theorem}\label{thm1}
Equation \eqref{eq3} admits a four-dimensional Lie algebra spanned by the following generators: 
\[
\mathcal{V}_1=\frac{\partial }{\partial t},~ 
\mathcal{V}_2=\frac{\partial }{\partial x},~ 
\mathcal{V}_3=t\frac{\partial }{\partial x} + \frac{\partial }{\partial u},~
\mathcal{V}_4=-2v\frac{\partial }{\partial v} + 2t\frac{\partial }{\partial t} 
- u\frac{\partial }{\partial u} + x\frac{\partial }{\partial x}.
\]
\end{theorem}
 Let's consider the  ordinary differential equations (ODEs) and initial conditions corresponding to each $\mathcal{V}_i$ ($i=1(1)4$) as 
\[
\frac{d\bar{x}}{d\epsilon} = \xi_1(\bar{x}, \bar{t}, \bar{u},  \bar{v}), \quad \bar{x}|_{\epsilon = 0} = x, \quad
\frac{d\bar{t}}{d\epsilon} = \xi_2(\bar{x}, \bar{t}, \bar{u},  \bar{v}), \quad \bar{t}|_{\epsilon = 0} = t,
\]
\[
\frac{d\bar{u}}{d\epsilon} = \eta_1(\bar{x}, \bar{t}, \bar{u},  \bar{v}), \quad \bar{u}|_{\epsilon = 0} = u,\quad
\frac{d\bar{v}}{d\epsilon} = \eta_2(\bar{x}, \bar{t}, \bar{u},  \bar{v}), \quad \bar{v}|_{\epsilon = 0} = v,
\]
which give some one-parameter groups \( \mathcal{G}_i \) associated with each \( \mathcal{V}_i \) as follows:  
\[
\mathcal{G}_1 : (x, t, u, v) \mapsto (x, t + \epsilon, u, v), \quad
\mathcal{G}_2 : (x, t, u, v) \mapsto (x + \epsilon, t, u, v),
\]
\[
\mathcal{G}_3 : (x, t, u, v) \mapsto 
\left(x+\epsilon t, t, u+\epsilon, v
\right), \quad \mathcal{G}_4 : (x, t, u, v) \mapsto 
\left(x e^{\epsilon}, t e^{2 \epsilon}, u e^{-\epsilon}, v e^{-2 \epsilon}
\right).
\]
This gives the following theorem.
\begin{theorem}\label{thm2}
The Lie symmetry generators $\mathcal{V}_1, \mathcal{V}_2, \mathcal{V}_3,$ $ \mathcal{V}_4$ for the system \eqref{eq3} produce the symmetry group of transformations,
 $\mathcal{G}_1 : (x, t, u, v) \mapsto (x, t + \epsilon, u, v),~
\mathcal{G}_2 : (x, t, u, v) \mapsto (x + \epsilon, t, u, v),~\mathcal{G}_3 : (x, t, u, v) \mapsto 
\left(x+\epsilon t, t, u+\epsilon, v
\right), ~ \mathcal{G}_4 : (x, t, u, v) \mapsto 
\left(x e^{\epsilon}, t e^{2 \epsilon}, u e^{-\epsilon}, v e^{-2 \epsilon}
\right)$, respectively, and the associated group invariant solutions 
 are, $u^{(1)}=\Lambda(x,t-\epsilon), ~v^{(1)}=\Gamma(x,t-\epsilon)$;~ 
 $u^{(2)}=\Lambda(x-\epsilon,t), ~v^{(2)}=\Gamma(x-\epsilon,t);$~$u^{(3)}=\Lambda(x-\epsilon t,t)-\epsilon, ~v^{(3)}=\Gamma(x-\epsilon t,t);$~$u^{(4)}=e^{\epsilon}\Lambda(x e^{\epsilon},t e^{-2 \epsilon}), ~v^{(4)}=e^{2 \epsilon}\Gamma(x e^{\epsilon},t e^{-2 \epsilon}).$
\end{theorem}

Additionally, applying the Lie bracket definition, i.e., $[\mathcal{V}_s, \mathcal{V}_t] = \mathcal{V}_s \mathcal{V}_t - \mathcal{V}_t \mathcal{V}_s$, the commutators read
\begin{align}\label{eq10}
\nonumber [\mathcal{V}_1, \mathcal{V}_1] = [\mathcal{V}_2, \mathcal{V}_2] = [\mathcal{V}_3, \mathcal{V}_3] = [\mathcal{V}_4, \mathcal{V}_4] = 0,
[\mathcal{V}_1, \mathcal{V}_2] = -[\mathcal{V}_2, \mathcal{V}_1] = 0, & \\ \nonumber 
[\mathcal{V}_1, \mathcal{V}_3] = -[\mathcal{V}_3, \mathcal{V}_1] = \mathcal{V}_2,
[\mathcal{V}_1, \mathcal{V}_4] = -[\mathcal{V}_4, \mathcal{V}_1] = 2\mathcal{V}_1,
[\mathcal{V}_2, \mathcal{V}_3] = -[\mathcal{V}_3, \mathcal{V}_2] = 0, & \\ 
[\mathcal{V}_2, \mathcal{V}_4] = -[\mathcal{V}_4, \mathcal{V}_2] = \mathcal{V}_2,
[\mathcal{V}_3, \mathcal{V}_4] = -[\mathcal{V}_4, \mathcal{V}_3] = -\mathcal{V}_3.
\end{align}

These commutation relations between $\mathcal{V}_i$'s yield the following theorem.
\begin{theorem}\label{thm3}
The set of generators $\mathcal{V}_i$ $(i = 1, 2, 3, 4)$ in Theorem \ref{thm1} forms a four-dimensional symmetry Lie algebra $\mathcal{L}_4$.
\end{theorem}
In the following, we utilize the Lie series representation $\text{Ad}(\exp(\epsilon \mathcal{V}_i)) \mathcal{V}_j = \mathcal{V}_j - \epsilon [\mathcal{V}_i, \mathcal{V}_j] 
+ \frac{1}{2} \epsilon^2 [\mathcal{V}_i, [\mathcal{V}_i, \mathcal{V}_j]] - \cdots \quad \text{for} \ \epsilon \in \mathcal{R}$ to derive the adjoint representations associated with the given vector fields.
 By using  \eqref{eq10} we have,
\[
\text{Ad}(\exp(\epsilon \mathcal{V}_i)) \mathcal{V}_i = \mathcal{V}_i, \quad i = 1, 2, 3, 4
\]
\[
\text{Ad}(\exp(\epsilon \mathcal{V}_1)) \mathcal{V}_2 = \mathcal{V}_2, \quad 
\text{Ad}(\exp(\epsilon \mathcal{V}_1)) \mathcal{V}_3 = \mathcal{V}_3 - \epsilon \mathcal{V}_2, \quad \text{Ad}(\exp(\epsilon \mathcal{V}_1)) \mathcal{V}_4 = \mathcal{V}_4 - 2 \epsilon \mathcal{V}_1,
\]
\[
\text{Ad}(\exp(\epsilon \mathcal{V}_2)) \mathcal{V}_1 = \mathcal{V}_1, \quad 
\text{Ad}(\exp(\epsilon \mathcal{V}_2)) \mathcal{V}_3 = \mathcal{V}_3, \quad \text{Ad}(\exp(\epsilon \mathcal{V}_2)) \mathcal{V}_4 = \mathcal{V}_4 - \epsilon \mathcal{V}_2,
\]
\[
\text{Ad}(\exp(\epsilon \mathcal{V}_3)) \mathcal{V}_1 = \mathcal{V}_1+\epsilon \mathcal{V}_2, \quad 
\text{Ad}(\exp(\epsilon \mathcal{V}_3)) \mathcal{V}_2 = \mathcal{V}_2, \quad \text{Ad}(\exp(\epsilon \mathcal{V}_3)) \mathcal{V}_4 = \mathcal{V}_4 + \epsilon \mathcal{V}_3,
\]
\begin{equation}\label{eq11}
\text{Ad}(\exp(\epsilon \mathcal{V}_4)) \mathcal{V}_1 = e^{2 \epsilon}\mathcal{V}_1, \quad 
\text{Ad}(\exp(\epsilon \mathcal{V}_4)) \mathcal{V}_2 = e^{\epsilon}\mathcal{V}_2, \quad \text{Ad}(\exp(\epsilon \mathcal{V}_4)) \mathcal{V}_3 = e^{-\epsilon}\mathcal{V}_3.
\end{equation}

Based on the adjoint representations given by \eqref{eq11}, the corresponding optimal set of subalgebras can be found from the following theorem.

\begin{theorem}
The one-dimensional optimal set of subalgebras corresponding to the Lie algebra \( \mathcal{L}_4 \) of the governing equations \eqref{eq3} is generated by the set of symmetry operators  
$$
\{ \mathcal{V}_1, \mathcal{V}_2, \mathcal{V}_3, \mathcal{V}_4, \mathcal{V}_1 + c \mathcal{V}_2,  \mathcal{V}_1 + c \mathcal{V}_3, \mathcal{V}_4 + c \mathcal{V}_3 \},
$$
where  $c \in \{-1, 1\}$ .
\end{theorem}
\section{Over view of the Extended Simplest Equation Method}\label{sec2}
The extended simplest equation method is a powerful analytical technique developed as an enhancement of the traditional simplest equation method \cite{ref32}, designed to construct exact traveling wave solutions of nonlinear evolution equations. The various important steps of this method for a general nonlinear partial differential equation are discussed as follows:

Let us consider a general nonlinear partial differential equation with independent variables $x, t$ and dependent variable $w(x,t)$, expressed as:
\begin{equation}\label{neq1}
F(w, w_x, w_t, w_{xx}, w_{xt}, w_{tt}, \ldots) = 0,
\end{equation}
where \( F \) is a polynomial function of its arguments, involving both nonlinear terms and the highest-order derivatives of $w(x,t)$.

To apply the extended simplest equation method to the above general nonlinear PDE \eqref{neq1} effectively, we proceed through the following main steps:

\textbf{Step 1.} Utilize the traveling wave transformation
\begin{equation}
w(x, t) = w(\zeta), \quad \zeta = kx - ct, 
\end{equation}
which converts the Eq. \eqref{neq1} into an ordinary differential equation (ODE) in terms of \( w(\zeta) \) as:
\begin{equation}\label{ode}
Q(w, kw', -cw', k^2w'', -kcw'', c^2w'', \ldots) = 0.
\end{equation}

\textbf{Step 2.} Identify the positive integer $M$ by balancing the highest-order derivative with the leading nonlinear term in Eq. \eqref{ode}.

\textbf{Step 3.} Let consider the solution to Eq. \eqref{ode} in the following  form:
\begin{equation}\label{sform}
w(\zeta) = \sum_{i_1=0}^{M} a_{i_1} \left( \frac{\psi'}{\psi} \right)^{i_1} + \sum_{i_2=0}^{M-1} b_{i_2} \left( \frac{\psi'}{\psi} \right)^{i_2} \frac{1}{\psi},
\end{equation}
where \( a_{i_1} \) and \( b_{i_2} \) \( (i_1 = 0, 1, \ldots, M;\; i_2 = 0, 1, \ldots, M-1) \) are constants, and \( a_M b_{M-1} \neq 0 \). Further, the function \( \psi = \psi(\zeta) \) satisfies the second-order linear ODE:
\begin{equation}\label{eq55}
\psi'' + d\psi = m, 
\end{equation}
with some constants \( d \) and \( m \). The equation \eqref{eq5} provides three distinct types of general solutions based on choosing the parameter $d$:
\begin{equation}\label{eq77}
\psi(\zeta) =
\begin{cases}
\mathcal{A}_1 \cosh(\sqrt{-d}\,\zeta) + \mathcal{A}_2 \sinh(\sqrt{-d}\,\zeta) + \dfrac{m}{d}, & \text{if } d < 0, \\
\mathcal{A}_1 \cos(\sqrt{d}\,\zeta) + \mathcal{A}_2 \sin(\sqrt{d}\,\zeta) + \dfrac{m}{d}, & \text{if } d > 0, \\
\dfrac{m}{2} \zeta^2 + \mathcal{A}_1 \zeta + \mathcal{A}_2, & \text{if } d = 0,
\end{cases}
\end{equation}

and
\begin{equation}\label{eq7}
\left( \frac{\psi'}{\psi} \right)^2 =
\begin{cases}
\dfrac{d\mathcal{A}_1^2 - d\mathcal{A}_2^2 - \dfrac{m^2}{d}}{\psi^2} - d + \dfrac{2m}{\psi}, & \text{if } d < 0, \\
\dfrac{d\mathcal{A}_1^2 + d\mathcal{A}_2^2 - \dfrac{m^2}{d}}{\psi^2} - d + \dfrac{2m}{\psi}, & \text{if } d > 0, \\
\dfrac{\mathcal{A}_1^2 - 2m\mathcal{A}_2}{\psi^2} + \dfrac{2m}{\psi}, & \text{if } d = 0,
\end{cases}
\end{equation}

where \( \mathcal{A}_1 \) and \( \mathcal{A}_2 \) are arbitrary constants.

\textbf{Step 4.} By substituting the expression \eqref{sform} into Eq.~\eqref{neq1}, and applying the second-order linear differential equation \eqref{eq55} together with the relations given in \eqref{eq7}, we collect terms according to powers of $\frac{1}{\psi}$ and $\frac{1}{\psi^i}\frac{\psi'}{\psi}$. This transforms the left-hand side of equation~\eqref{neq1} into a polynomial in $\frac{1}{\psi}$ and $\frac{1}{\psi^i}\frac{\psi'}{\psi}$. Equating the coefficients of each distinct power to zero yields a system of nonlinear algebraic equations involving the parameters $a_{i_1}, b_{i_2} \ (i_1 = 0, 1, \ldots, M; \ i_2 = 0, 1, \ldots, M - 1), \ k, \ c, \ d$, and $m$.

\textbf{Step 5.} Assume that the constants \( a_{i_1}, b_{i_2} \ (i_1 = 0, 1, \ldots, M;\ i_2 = 0, 1, \ldots, M - 1),\ k,\ c,\ d \), and \( m \) are determined by solving the system of nonlinear algebraic equations obtained in Step~4. Substituting these values, along with the general solution \eqref{sform} of the differential equation \eqref{ode}, into Eq.~\eqref{neq1}, enables the construction of more precise traveling wave solutions.

\textbf{Remark:} Notably, more exact solutions of Eq.~\eqref{neq1} are achievable when $a_M b_{M-1} = 0$.


\section {Symmetry reductions and exact solutions}\label{sec3}
In this section, we explore the symmetry reductions and exact solutions of the governing system \eqref{eq3} corresponding to each optimal subalgebra. 

{\bf (I) For the generator $\mathcal{V}_1:$}   

The characteristic equation for the generator $\mathcal{V}_1$ gives the solution ansatzs in the form $u = F(x), v=G(x)$. By substituting these ansatzs into \eqref{eq3}, we have the following set of ordinary differential equations:
\begin{equation}
F(x) F'(x) + G'(x) + \alpha F''(x) = 0, \quad 
F(x) G'(x) + G(x) F'(x) + \tau F'''(x) - \alpha G''(x) = 0.
\end{equation}
Thus, applying Lie symmetry analysis transforms the complex system of nonlinear partial differential equations into a simpler non-autonomous system of ordinary differential equations, making it relatively easier to solve. Moreover, the non-autonomous system can be further converted into a more tractable autonomous system for $F(x)$ as follows:

\begin{equation}\label{equa1}
(\tau+\alpha^2) F'''(x)-\frac{3}{2}F'(x) F(x)^2=0,
\end{equation}
whereas $G(x)$ can be determined by the expression $G(x)=-\alpha F'(x)-\frac{F(x)^2}{2}$.
Solving Eq.~\eqref{equa1} by using the extended simplest equation method, we obtain three distinct sets of solutions in the general form \[
F(x) = a_0 + a_1 \frac{\psi'(x)}{\psi(x)} + \frac{a_1}{\psi(x)},
\] that are not available in the existing literature. Here, the unknown function $\psi(x)$ and unknown coefficients $a_0, a_1, a_2$ are determined for each cases as follows: 

\textbf{Case-i:} This case provides the mixture of hyperbolic  and rational solutions in the following form,
\[
\psi(x) = \upsilon_1 \cosh(x) + \upsilon_2 \sinh(x) - p, \quad \text{with } \upsilon_1, \upsilon_2 \in \mathcal{R}, \upsilon_2^2 > \upsilon_1^2, \alpha^2+\tau > 0.
\]
The possible parameter choices  for this case are as follows:

\begin{align*}
     (a_0, a_1, a_2, p) \in \Bigg\{ &\left(\frac{1}{3} \sqrt{6\alpha^2 + 6\tau},~ 0,~   \frac{6p(\alpha^2 + \tau)}{\sqrt{6\alpha^2 + 6\tau}},   
     ~\pm  \sqrt{-2\upsilon_1^2 + 2\upsilon_2^2}\right), \\  
     &\left(-\frac{1}{3} \sqrt{6\alpha^2 + 6\tau},~ 0,~   -\frac{6p(\alpha^2 + \tau)}{\sqrt{6\alpha^2 + 6\tau}},   
     ~\pm  \sqrt{-2\upsilon_1^2 + 2\upsilon_2^2}\right) \Bigg\}.
\end{align*}

\textbf{Case-ii:} In this case,
\[
\psi(x) = \upsilon_1 \cos(x) + \upsilon_2 \sin(x) + p, \quad \text{with } \alpha^2 + \tau < 0.
\]
 For this case, the suitable choices for the parameter values are as follows:

\begin{itemize}
    \item \( a_0 = \frac{1}{3} \sqrt{-6\alpha^2 - 6\tau}, \quad  
    a_1 = 0, \quad  
    a_2 = \frac{6p(\alpha^2 + \tau)}{\sqrt{-6\alpha^2 - 6\tau}}, \quad  
    p = \pm i \sqrt{2\upsilon_1^2 + 2\upsilon_2^2}, \quad i=\sqrt{-1} \).
    
    \item Alternatively,  
    \( a_0 = -\frac{1}{3} \sqrt{-6\alpha^2 - 6\tau}, \quad  
    a_1 = 0, \quad  
    a_2 = -\frac{6p(\alpha^2 + \tau)}{\sqrt{-6\alpha^2 - 6\tau}}, \quad  
    p = \pm i \sqrt{2\upsilon_1^2 + 2\upsilon_2^2} \).
\end{itemize}
This case gives a complex velocity function in terms of trigonometric functions. In wave theory, a complex phase velocity implies that the wave is experiencing attenuation (if the imaginary part is negative) or amplification (if positive). In general, the imaginary component of velocity often provides insight into non-conservative effects such as dissipation, instability, or phase behavior.

\textbf{Case-iii:} Here,
\[
\psi(x) = \frac{p}{2}x^2+\upsilon_1 x+\upsilon_2 ~\text{with}~ \alpha^2+\tau \geq 0, p, \upsilon_1, \upsilon_2 \in \mathcal{R}.
\]
The parameters are determined as follows:

\begin{itemize}
    \item \( a_0 = 0, \quad  
    a_1 = \sqrt{(\alpha^2+\tau)}, \quad  
    a_2 = \sqrt{\upsilon_1^2 \alpha^2 - 2 \upsilon_2 \alpha^2 p + \upsilon_1^2 \tau - 2 \upsilon_2 p \tau} \) with $\upsilon_1^2 \alpha^2 - 2 \upsilon_2 \alpha^2 p + \upsilon_1^2 \tau - 2 \upsilon_2 p \tau \geq 0$.
    
    \item Alternatively,  
    \( a_0 =0, \quad  
    a_1 = -\sqrt{(\alpha^2+\tau)}, \quad  
    a_2 = -\sqrt{\upsilon_1^2 \alpha^2 - 2 \upsilon_2 \alpha^2 p + \upsilon_1^2 \tau - 2 \upsilon_2 p \tau} \) with $\upsilon_1^2 \alpha^2 - 2 \upsilon_2 \alpha^2 p + \upsilon_1^2 \tau - 2 \upsilon_2 p \tau \geq 0$.
\end{itemize}
Hence, from these cases, we obtain a new class of hyperbolic, trigonometric, and polynomial-type solutions for WBK equations, which were not achievable from the past studies.\\\\

\textbf{(II) For the generator $\mathcal{V}_2:$}

The characteristic equation for the generator $\mathcal{V}_2$ yields the solution ansatzs $u = F(t), v=G(t)$. Substitution of these ansatzs into \eqref{eq3} yields $F'(t)=G'(t)=0,$ which gives the trivial solutions $u(x,t)=c_{21}, v(x,t)=c_{22}$. Here, $c_{21}, c_{22}$ are arbitrary constants.\\\\

\textbf{(III) For the generator $\mathcal{V}_3:$}

The characteristic equation for the generator $\mathcal{V}_3$, gives $u = \frac{x}{t}+F(t), v=G(t)$. Substitution of these ansatzs into \eqref{eq3} yields 
\begin{equation}
{F'(t)t + F(t)}= 0, \quad t G'(t) + G(t) = 0.
\end{equation}
Hence, the final form of solutions for this case are $u(x,t)=\frac{x}{t} + \frac{c_{31}}{t}, \; v(x,t)=\frac{c_{32}}{t}$, where $c_{31}, c_{32}$ are arbitrary constants.\\\\

\textbf{ (IV) For the generator $\mathcal{V}_4:$ }  

The characteristic equation for the generator $\mathcal{V}_4$, yields the similarity transformations $u = \frac{1}{x}F(\xi), v=\frac{1}{t}G(\xi)$, where $\xi= \frac{x}{\sqrt{t}}$. Substitution of these ansatzs into \eqref{eq3} yields 
\begin{equation}\label{eq20251}
\begin{aligned}
& 2 F''(\xi) \alpha \xi^2 - F'(\xi) \xi^3 + 2 G'(\xi) \xi^3 + 2 F'(\xi) F(\xi) \xi - 4 F'(\xi) \alpha \xi - 2 F(\xi)^2 + 4 F(\xi) \alpha = 0, \\
& G'(\xi) \xi^5 + 2 G''(\xi) \alpha \xi^4 - 2 F'(\xi) G(\xi) \xi^3 - 2 F(\xi) G'(\xi) \xi^3 + 2 G(\xi) \xi^4 - 2 F'''(\xi) \xi^3 \tau  + 6 F''(\xi) \xi^2 \tau \\ & + 2 F(\xi) G(\xi) \xi^2 - 12 F'(\xi) \xi \tau + 12 F(\xi) \tau = 0.
\end{aligned}
\end{equation}
This shows that the similarity transformation obtained through symmetry analysis simplifies the complex system of partial differential equations by reducing it to a set of ordinary differential equations. The polynomial solution for the resulting system \eqref{eq20251} is given by $F(\xi) = \frac{2 \xi^2}{3}$ and $G(\xi) = -\frac{2 \alpha}{3} + \frac{\xi^2}{9}$. Furthermore, the reduced system \eqref{eq20251} can be solved numerically under suitable boundary conditions using methods such as the shooting technique or finite difference schemes.\\\\

\textbf{(V) For the generator $\mathcal{V}_1+c \mathcal{V}_3:$}

The characteristic equation for the generator $\mathcal{V}_1+c \mathcal{V}_3$, yields $u = t+F(\xi), v=G(\xi)$, where $\xi= \frac{t^2}{2}-x$. Substitution of these ansatzs into \eqref{eq3} results the following 
\begin{equation}\label{equa2}
\begin{aligned}
& - F'(\xi) F(\xi) + \alpha F''(\xi) - G'(\xi) + 1 = 0, \\
&  G(\xi) F'(\xi) + \tau F'''(\xi) + \alpha G''(\xi) +G'(\xi) F(\xi) = 0.
\end{aligned}
\end{equation}
The set of two independent solutions for the above system \eqref{equa2} for $\alpha=\tau=0$ is given by, $F(\xi)=0, G(\xi)=\xi+c_{51}$ and 
\begin{align*}
F(\xi) &= \frac{1}{3} \left(-27 c_{51} + 3 \sqrt{24 c_{52}^3 + 72 c_{52}^2 \xi + 72 c_{52} \xi^2 + 24 \xi^3 + 81 c_{51}^2} \right)^{1/3} \\
&\quad - \frac{3 \left(\frac{2}{3} c_{52} + \frac{2}{3} \xi \right)}
{\left(-27 c_{51} + 3 \sqrt{24 c_{52}^3 + 72 c_{52}^2 \xi + 72 c_{52} \xi^2 + 24 \xi^3 + 81 c_{51}^2} \right)^{1/3}}
\end{align*}
while $G(\xi)$ can be obtained by the relation 
\begin{align*}
G(\xi) & = -\frac{F'(\xi) F(\xi)^2 + F(\xi)}{F'(\xi)} ~\mbox{with} F'(\xi), \left(-27 c_{51} + 3 \sqrt{24 c_{52}^3 + 72 c_{52}^2 \xi + 72 c_{52} \xi^2 + 24 \xi^3 + 81 c_{51}^2} \right) \neq 0, \mbox{and} \\ & 24 c_{52}^3 + 72 c_{52}^2 \xi + 72 c_{52} \xi^2 + 24 \xi^3 + 81 c_{51}^2\geq 0.
\end{align*}
Here, $c_{51}, c_{52}$ are arbitrary constants.\\
Further, the complicated non-autonomous  ODEs \eqref{equa2} can be converted into a single ordinary differential equation in terms of $F(\xi)$ as:
\begin{equation}\label{eq22}
    (\tau+\alpha^2)F''(\xi)-\frac{F(\xi)^3}{2}+\xi F(\xi)=0.
\end{equation}
For non-zero $\alpha$ and $\tau$ with $\alpha^2+\tau \neq 0$, the power series solution of the above system of equations \eqref{equa2} is given by
\begin{align*}
 F(\xi) = & c_{53} \left( 
1 + \frac{c_{53}^2}{2(\tau + \alpha^2)} \xi^2 
- \frac{1}{6(\tau + \alpha^2)} \xi^3 
+ \frac{3 c_{53}^4}{12 (\tau+\alpha^2)^2} \xi^4 + \frac{\frac{3}{2} c_{53}^4 + \frac{1}{2} c_{53}^2 - \frac{c_{53}^3}{2}}{20 (\tau+\alpha^2)} \xi^5 
+ \cdots 
\right) \\
& + c_{54} \left( 
\xi 
+ \frac{3 c_{53}^2}{6(\tau + \alpha^2)} \xi^3 
+ \frac{3 c_{54}^2 + 1}{12(\tau + \alpha^2)} \xi^4 
+ \frac{\frac{3}{2} c_{53}^4 + c_{54}^3}{20 (\tau + \alpha^2)} \xi^5 
+ \cdots 
\right),
\end{align*}
where $c_{53}, c_{54}$ are arbitrary constants.
Moreover, using this $F(\xi)$, one can determine the power series solution for $G(\xi)$ as $G(\xi)=\xi+\alpha F'(\xi)-\frac{F(\xi)^2}{2}.$\\\\



\textbf{(VI) For the generator $\mathcal{V}_3+c \mathcal{V}_4:$ } 

The characteristic equation for the generator $\mathcal{V}_3+c \mathcal{V}_4$, yields $u = 1+\frac{1}{\sqrt{t}}F(\xi), v=\frac{1}{t}G(\xi)$, where $\xi= \frac{x-t}{\sqrt{t}}$. Substitution of these ansatzs into \eqref{eq3} yields 
\begin{equation}\label{eq20252}
\begin{aligned}
& 2 F'( \xi ) F( \xi ) - F'( \xi ) \xi + 2 F''( \xi ) \alpha - F( \xi ) + 2 G'( \xi ) = 0, \\
& 2 G( \xi ) F'( \xi ) + 2 F'''( \xi ) \tau - 2 G''( \xi ) \alpha + 2 G'( \xi ) F( \xi ) - G'( \xi ) \xi - 2 G( \xi ) = 0.
\end{aligned}
\end{equation}
The simpler polynomial form solutions for the above system of ODEs \eqref{eq20252} is $F(\xi)=\frac{2 \xi}{3},$ $G(\xi)=\frac{-4 \alpha}{3}+\frac{2 \xi^2}{9}.$ This system can also be solved easily by using numerical methods (Finite difference, Shooting, R-K methods, etc.) under certain boundary conditions. \\\\


\textbf{ (VII) For the generator $\mathcal{V}_1+c \mathcal{V}_2:$ }

The characteristic equation for the generator $\mathcal{V}_1+c \mathcal{V}_2$, yields $u =F(\xi), v=G(\xi)$, where $\xi= x-c t$. Substitution of these ansatzs into \eqref{eq3} gives
\begin{equation}
\begin{aligned}\label{goveq2025}
& - F'( \xi ) c + F( \xi ) F'( \xi ) + G'( \xi ) + \alpha F''( \xi ) = 0, \\
& - G'( \xi ) c + F( \xi ) G'( \xi ) + G( \xi ) F'( \xi ) + \tau F'''( \xi ) - \alpha G''( \xi ) = 0.
\end{aligned}
\end{equation}
This system of equations can be solved using the extended simplest equation method. The obtained solutions represent exact traveling wave solutions, which can be expressed in terms of hyperbolic, trigonometric, and rational functions, respectively. Notably, the parameter $c$ represents the wave speed. The general form of the solution for the three different cases is $$
F(\xi) = a_0 + a_1 \frac{\psi'(\xi)}{\psi(\xi)} + a_2 \frac{1}{\psi(\xi)}, $$
$$
G(\xi) = b_0 + b_1 \frac{\psi'(\xi)}{\psi(\xi)} + b_2 \frac{\psi'(\xi)^2}{\psi(\xi)^2} + b_3 \frac{1}{\psi(\xi)} + b_4 \frac{1}{\psi(\xi)} \cdot \frac{\psi'(\xi)}{\psi(\xi)},$$ where the unknown function $\psi(\xi)$ and the unknown coefficients are given by following cases:  

\textbf{Case-i:} Here,
\[
\psi(\xi) = \upsilon_1 \cosh(\xi) + \upsilon_2 \sinh(\xi) - p, \quad \text{with } \upsilon_1, \upsilon_2 \in \mathcal{R}, \upsilon_2^2- \upsilon_1^2+p^2 > 0, \alpha^2+\tau > 0.
\]
This case provides the hyperbolic and rational solutions for WBK equations.
The possible parameter values for this case are determined as follows:
\begin{align*}
   & a_0 = \frac{\mathcal{M}_1}{(\alpha^2+\tau) (\upsilon_2^2-\upsilon_1^2+p^2)}, 
    a_1 = 0, 
    a_2 = 2 \sqrt{(\alpha^2+\tau) (\upsilon_2^2-\upsilon_1^2+p^2)}, \\ &
    b_0 = \frac{(\alpha^2+\tau)(\upsilon_1^2-\upsilon_2^2-2 p^2)}{\upsilon_1^2 - \upsilon_2^2 - p^2}, b_1=0, 
    b_2 = \frac{1}{2} \frac{a_2^2}{\upsilon_1^2 - \upsilon_2^2 - p^2},\\ & 
    b_3 = -\frac{a_2 (\upsilon_1^2 a_0 - \upsilon_1^2 c - \upsilon_2^2 a_0 + \upsilon_2^2 c - a_0 p^2 + c p^2 + a_2 p)}{\upsilon_1^2 - \upsilon_2^2 - p^2}, 
    b_4 = \alpha a_2.
\end{align*}
where,

$\mathcal{M}_1= c (\alpha^2+\tau) (\upsilon_2^2-\upsilon_1^2+p^2)  + (\alpha^2 p+p \tau) \sqrt{(\alpha^2+\tau) (\upsilon_2^2-\upsilon_1^2+p^2)}$.\\
 Alternatively,  
 \begin{align*}
   & a_0 = -\frac{\mathcal{M}_2}{(\alpha^2+\tau) (\upsilon_2^2-\upsilon_1^2+p^2)}, 
    a_1 = 0, 
    a_2 = -2 \sqrt{(\alpha^2+\tau) (\upsilon_2^2-\upsilon_1^2+p^2)}, \\ &
    b_0 = \frac{(\alpha^2+\tau)(\upsilon_1^2-\upsilon_2^2-2 p^2)}{\upsilon_1^2 - \upsilon_2^2 - p^2}, b_1=0, 
    b_2 = \frac{1}{2} \frac{a_2^2}{\upsilon_1^2 - \upsilon_2^2 - p^2},\\ & 
    b_3 = -\frac{a_2 (\upsilon_1^2 a_0 - \upsilon_1^2 c - \upsilon_2^2 a_0 + \upsilon_2^2 c - a_0 p^2 + c p^2 + a_2 p)}{\upsilon_1^2 - \upsilon_2^2 - p^2}, 
    b_4 = \alpha a_2.
\end{align*}
where,

$\mathcal{M}_2= c((\alpha^2+\tau) (\upsilon_2^2-\upsilon_1^2+p^2))  - (\alpha^2 p+p \tau) \sqrt{(\alpha^2+\tau) (\upsilon_2^2-\upsilon_1^2+p^2)}$.

\begin{figure}[t]
    \centering
    \begin{minipage}[b]{0.25\textwidth}
        \centering
        \includegraphics[width=\linewidth]{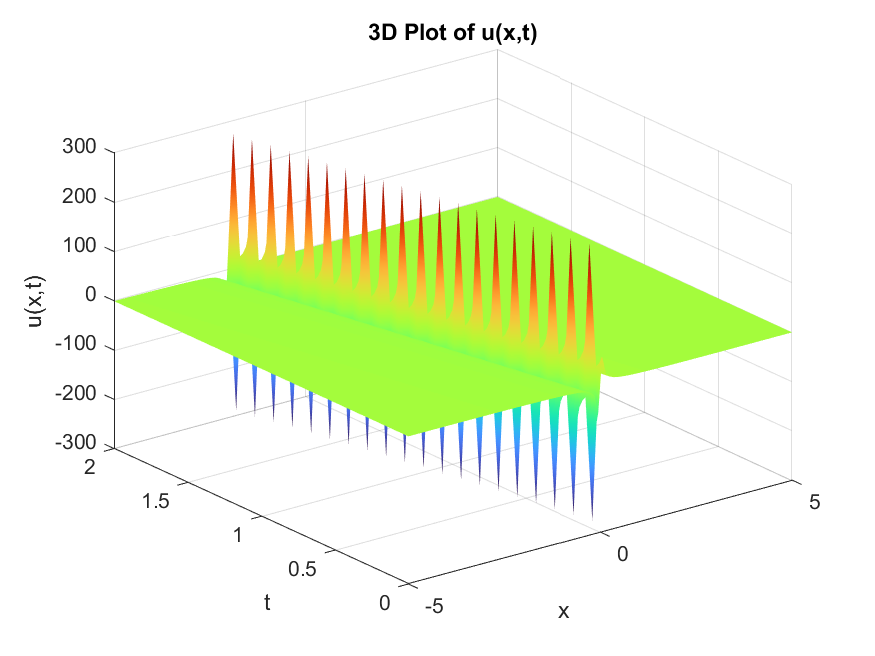}
        \caption*{(a) Plot for $c=-1$}
    \end{minipage}
    \hspace{0.01\textwidth}
    \begin{minipage}[b]{0.25\textwidth}
        \centering
        \includegraphics[width=\linewidth]{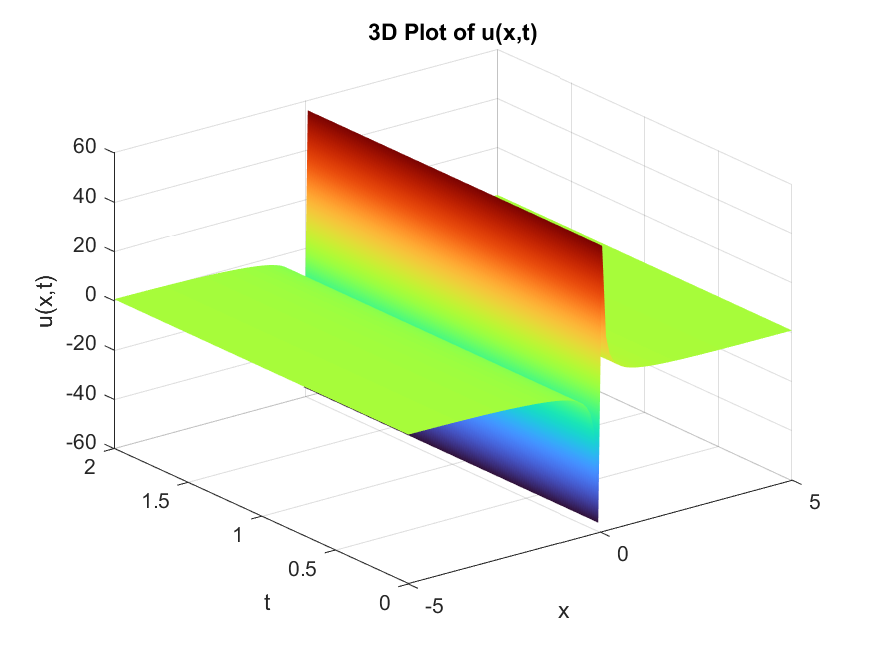}
        \caption*{(b) Plot for $c=0$}
    \end{minipage}
    \hspace{0.01\textwidth}
    \begin{minipage}[b]{0.25\textwidth}
        \centering
        \includegraphics[width=\linewidth]{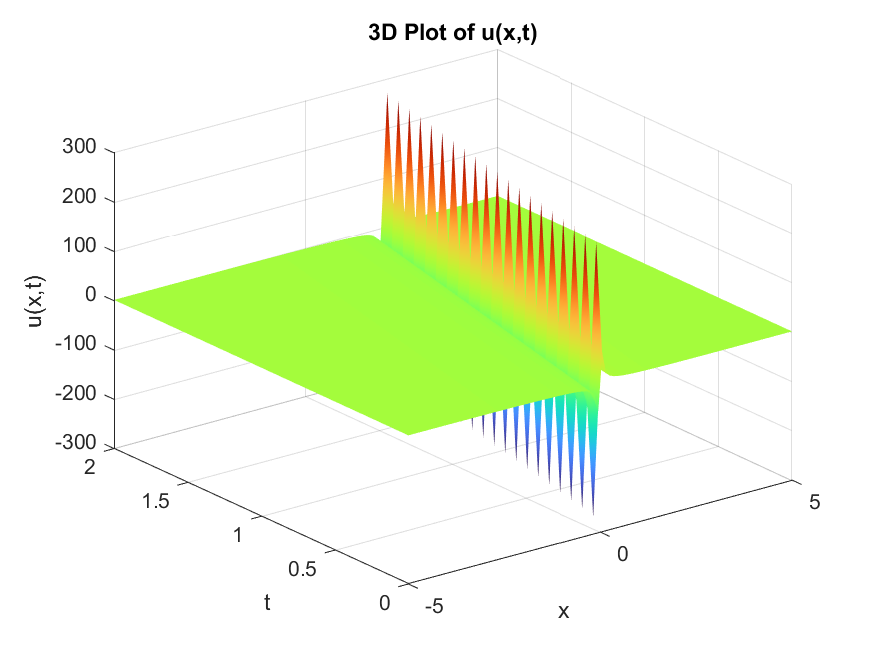}
        \caption*{(c) Plot for $c=1$}
    \end{minipage}
    \caption{Effect of the wave speed $c$ on the hyperbolic solution for $u(x,t)$, whenever the other parameter are fixed at $p=1, \alpha=1, \tau=1, \upsilon_1=1, \upsilon_2=2$.}
    \label{fig:1}
\end{figure}

Figure~\ref{fig:1} illustrates the effect of wave speed $c$ on the hyperbolic solution $u(x,t)$. As $c$ varies from $-1$ to $1$, the solution transitions from steep oscillatory behavior to a flatter, more stable wave profile. These results emphasize the critical role of wave speed in shaping the solution dynamics, shedding light on the influence of propagation velocity in nonlinear wave models.


\textbf{Case-ii:} In this case, we obtain trigonometric solutions in the form
\[
\psi(x) = \upsilon_1 \cos(\xi) + \upsilon_2 \sin(\xi) + p, \quad \text{with } \upsilon_1, \upsilon_2, p \in \mathcal{R}, \upsilon_1^2+ \upsilon_2^2-p^2 > 0,~ \alpha^2+\tau > 0.
\]
The suitable choices of parameter values are as follows:
\begin{align*}
   & a_0 = -\frac{\mathcal{M}_3}{(\alpha^2+\tau) (\upsilon_1^2  + \upsilon_2^2  - p^2) }, 
    a_1 = 0, a_2 = 2 \sqrt{(\alpha^2+\tau) (\upsilon_1^2+\upsilon_1^2-p^2)} \\ &
b_0 = \frac{(\alpha^2+\tau)(2 p^2-\upsilon_1^2-\upsilon_2^2)}{\upsilon_1^2 +\upsilon_2^2 - p^2},  
b_1=0,
b_2 = -\frac{1}{2} \frac{a_2^2}{\upsilon_1^2 + \upsilon_2^2 - p^2},\\ & 
    b_3 = -\frac{a_2 (\upsilon_1^2 a_0 - \upsilon_1^2 c + \upsilon_2^2 a_0 - \upsilon_2^2 c - a_0 p^2 + c p^2 - a_2 p)}{\upsilon_1^2 + \upsilon_2^2 - p^2}, 
    b_4 = \alpha a_2.
\end{align*}
where,

$\mathcal{M}_3= c (\alpha^2+\tau) (\upsilon_1^2+\upsilon_2^2-p^2)  + (\alpha^2 p+p \tau) \sqrt{(\alpha^2+\tau) (\upsilon_1^2+\upsilon_2^2-p^2)}$.\\
 Alternatively,  
 \begin{align*}
   & a_0 = -\frac{\mathcal{M}_4}{(\alpha^2+\tau) (\upsilon_1^2  + \upsilon_2^2  - p^2) }, 
    a_1 = 0, a_2 =- 2 \sqrt{(\alpha^2+\tau) (\upsilon_1^2+\upsilon_1^2-p^2)} \\ &
b_0 = \frac{(\alpha^2+\tau)(2 p^2-\upsilon_1^2-\upsilon_2^2)}{\upsilon_1^2 +\upsilon_2^2 - p^2},  
b_1=0,
b_2 = -\frac{1}{2} \frac{a_2^2}{\upsilon_1^2 + \upsilon_2^2 - p^2},\\ & 
    b_3 = -\frac{a_2 (\upsilon_1^2 a_0 - \upsilon_1^2 c + \upsilon_2^2 a_0 - \upsilon_2^2 c - a_0 p^2 + c p^2 - a_2 p)}{\upsilon_1^2 + \upsilon_2^2 - p^2}, 
    b_4 = \alpha a_2.
\end{align*}
where,

$\mathcal{M}_4= c (\alpha^2+\tau) (\upsilon_1^2+\upsilon_2^2-p^2)  - (\alpha^2 p+p \tau) \sqrt{(\alpha^2+\tau) (\upsilon_1^2+\upsilon_2^2-p^2)}$.\\

\begin{figure}[t]
    \centering
    \begin{minipage}[b]{0.25\textwidth}
        \centering
        \includegraphics[width=\linewidth]{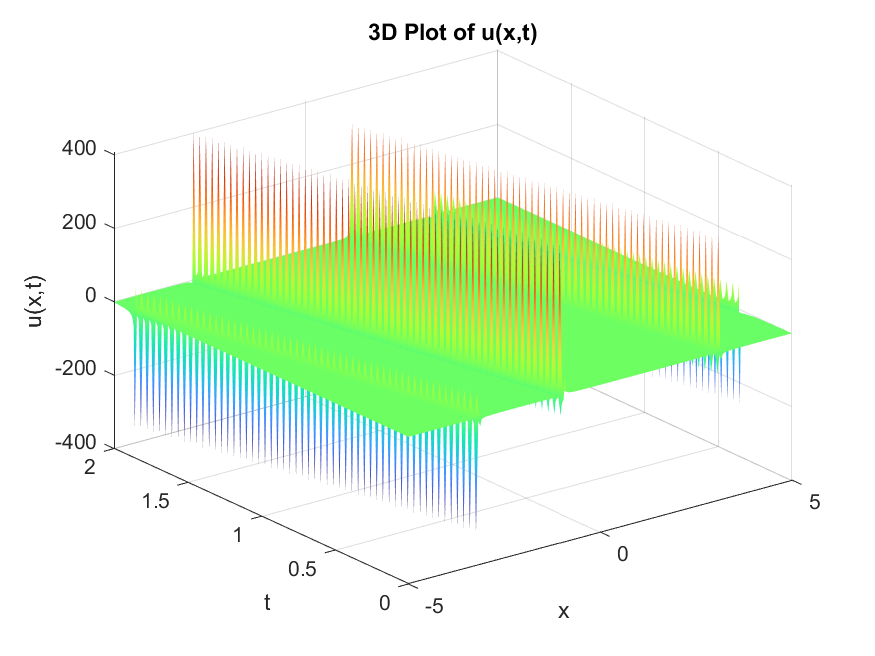}
        \caption*{(a) Plot for $c=-1$}
    \end{minipage}
    \hspace{0.01\textwidth}
    \begin{minipage}[b]{0.25\textwidth}
        \centering
        \includegraphics[width=\linewidth]{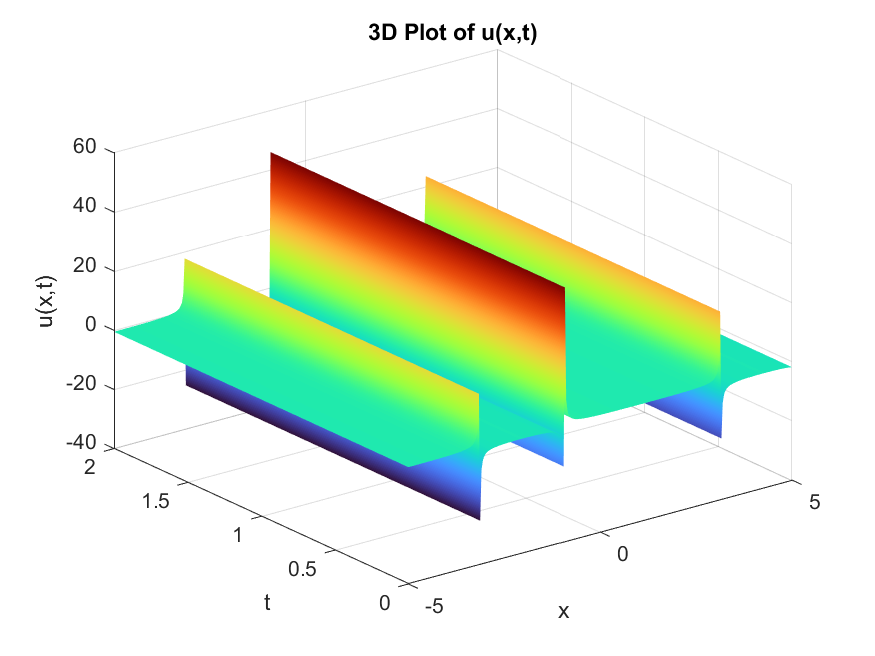}
        \caption*{(b) Plot for $c=0$}
    \end{minipage}
    \hspace{0.01\textwidth}
    \begin{minipage}[b]{0.25\textwidth}
        \centering
        \includegraphics[width=\linewidth]{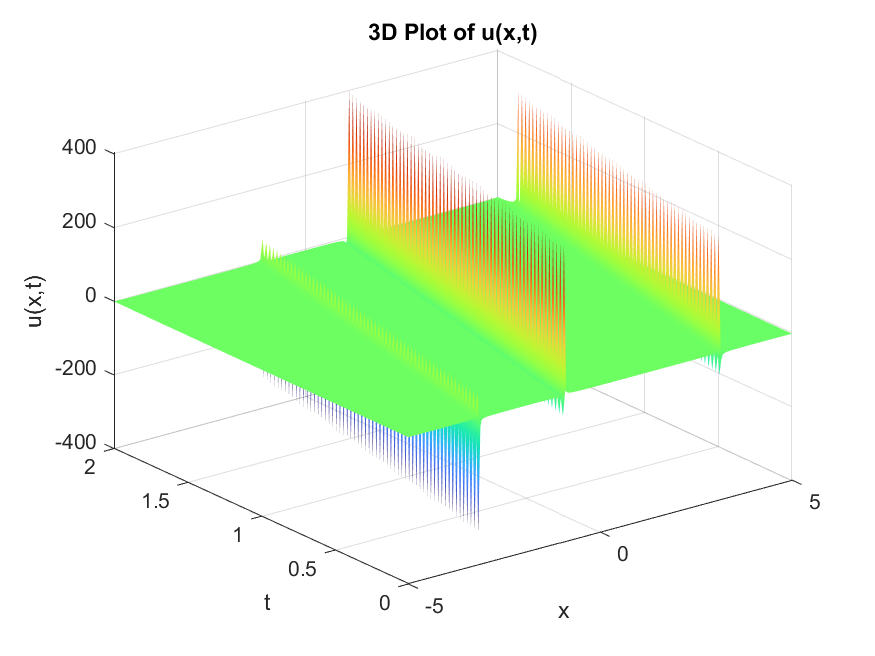}
        \caption*{(c) Plot for $c=1$}
    \end{minipage}
    \caption{Effect of the wave speed $c$ on the trigonometric solution for $u(x,t)$, whenever the other parameter are fixed at $p=1, \alpha=1, \tau=1, \upsilon_1=1, \upsilon_2=2$.}
    \label{fig:2}
\end{figure}

Figure~\ref{fig:2} demonstrates how the wave speed $c$ influences the trigonometric solution for $u(x,t)$ of the system. For $c = -1$ and $c = 1$, the solution surfaces exhibit sharp peaks and discontinuities, signifying wave interactions. In contrast, the solution for $c = 0$ remains symmetric and less steep, indicating a stationary or balanced wave behavior.


\textbf{Case-iii:} This case gives the rational solutions in the form
\[
\psi(x) = \frac{p}{2}\xi^2+\upsilon_1 \xi+\upsilon_2 ~\text{with}~ \alpha^2+\tau > 0, p, \upsilon_1, \upsilon_2 \in \mathcal{R} ~\text{with}~ \upsilon_1^2-2 \upsilon_2^2 p>0,
\]
and the unknown parameters are determined as follows:
\begin{align*}
   & a_0 = \frac{\alpha^2 p + c \sqrt{(\alpha^2 + \tau)(\upsilon_1^2 - 2 \upsilon_2 p)}  + \tau p}{\sqrt{(\alpha^2 + \tau)(\upsilon_1^2 - 2 \upsilon_2 p)}},
    a_1 = 0,  a_2 = 2 \sqrt{\upsilon_1^2 \alpha^2 - 2 \upsilon_2 \alpha^2 p + \upsilon_1^2 \tau - 2 \upsilon_2 p \tau}, \\ &
    b_0 = \frac{(\alpha^2 + \tau) p^2}{\upsilon_1^2 - 2 \upsilon_2 p}, 
    b_1 = 0, 
    b_2 = -\frac{1}{2} \frac{a_2^2}{\upsilon_1^2 - 2 \upsilon_2 p},
    b_3 = -\frac{a_2 (\upsilon_1^2 a_0 - \upsilon_1^2 c - 2 \upsilon_2 a_0 p + 2 \upsilon_2 c p - a_2 p)}{\upsilon_1^2 - 2 \upsilon_2 p}, 
    b_4 = \alpha a_2.
\end{align*}

Alternatively,

\begin{align*}
   & a_0 = \frac{-\alpha^2 p + c \sqrt{(\alpha^2 + \tau)(\upsilon_1^2 - 2 \upsilon_2 p)} - \tau p}{\sqrt{(\alpha^2 + \tau)(\upsilon_1^2 - 2 \upsilon_2 p)}},
    a_1 = 0,  a_2 = -2 \sqrt{\upsilon_1^2 \alpha^2 - 2 \upsilon_2 \alpha^2 p + \upsilon_1^2 \tau - 2 \upsilon_2 p \tau}, \\ &
    b_0 = \frac{(\alpha^2 + \tau) p^2}{\upsilon_1^2 - 2 \upsilon_2 p}, 
    b_1 = 0, 
    b_2 = -\frac{1}{2} \frac{a_2^2}{\upsilon_1^2 - 2 \upsilon_2 p},
    b_3 = -\frac{a_2 (\upsilon_1^2 a_0 - \upsilon_1^2 c - 2 \upsilon_2 a_0 p + 2 \upsilon_2 c p - a_2 p)}{\upsilon_1^2 - 2 \upsilon_2 p}, 
    b_4 = \alpha a_2.
\end{align*}

\begin{figure}[htbp]
    \centering
    \begin{minipage}[b]{0.25\textwidth}
        \centering
        \includegraphics[width=\linewidth]{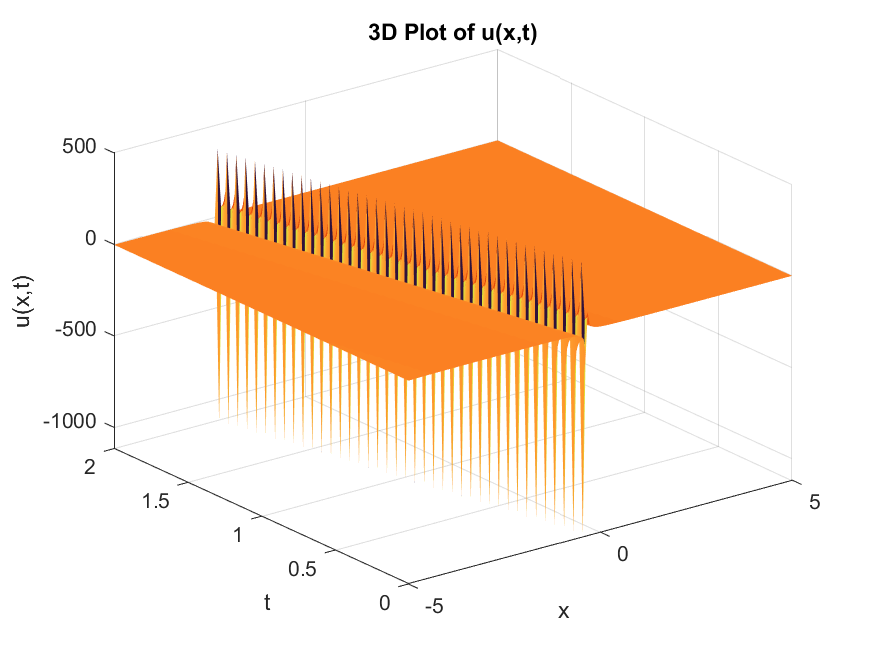}
        \caption*{(a) Plot for $c=-1$}
    \end{minipage}
    \hspace{0.01\textwidth}
    \begin{minipage}[b]{0.25\textwidth}
        \centering
        \includegraphics[width=\linewidth]{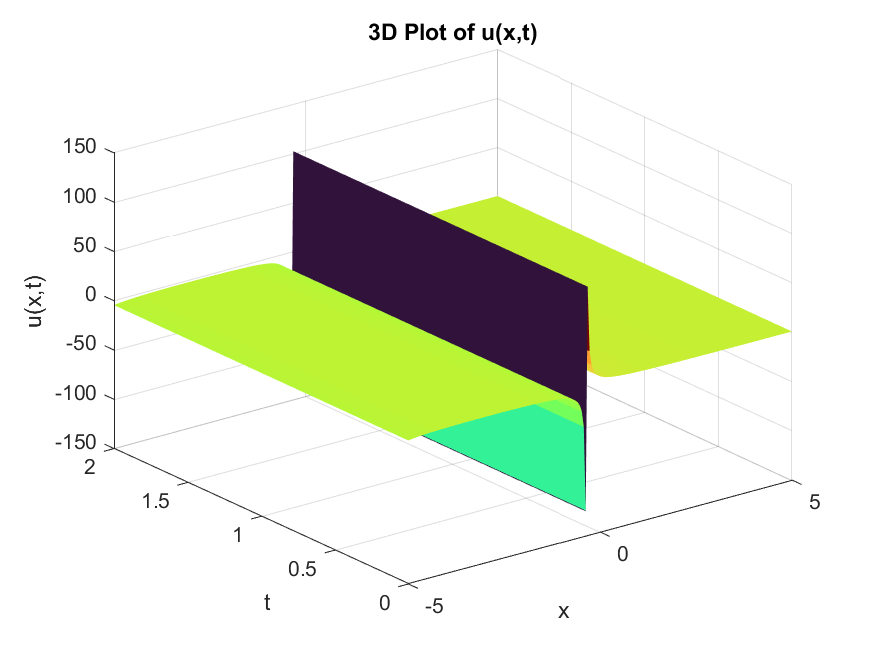}
        \caption*{(b) Plot for $c=0$}
    \end{minipage}
    \hspace{0.01\textwidth}
    \begin{minipage}[b]{0.25\textwidth}
        \centering
        \includegraphics[width=\linewidth]{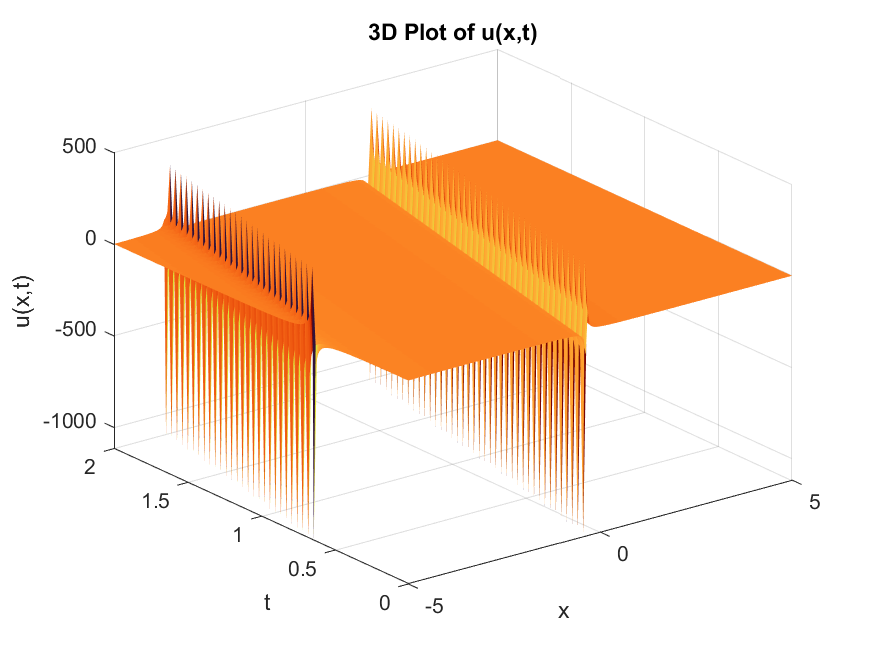}
        \caption*{(c) Plot for $c=1$}
    \end{minipage}
    \caption{Effect of the wave speed $c$ on the rational solution for $u(x,t)$, whenever the other parameter are fixed at $p=1, \alpha=1, \tau=1, \upsilon_1=1, \upsilon_2=2$.}
    \label{fig:3}
\end{figure}

Figure \ref{fig:3} illustrates the impact of varying wave speed $c$ on the rational solution for $u(x,t)$. For $c=-1$ and $c=1$, the solution exhibits steep wave fronts and sharp transitions, indicating dynamic wave propagation. In contrast, the case $c=0$ produces a stationary solution, reflecting the absence of wave movement.

Hence, from these three cases, we are getting some new kinds of hyperbolic, trigonometric, and rational type solutions. These cases also show that using symmetry analysis, one can reduce a complicated nonlinear system of partial differential equations into a simpler system of nonlinear ODEs, which can be either solved analytically using existing methods or numerically using suitable valid boundary conditions.

We now aim to numerically investigate the reduced system~\eqref{goveq2025} in order to explore traveling wave solutions of the Whitham--Broer--Kaup (WBK) system, under a set of physically meaningful boundary conditions. Our objective is to illustrate the influence of dispersive and nonlinear effects on the flow dynamics. As a first step, we aim to eliminate the auxiliary variable \( G(\xi) \) by differentiating the first equation and substituting it into the second, thereby reducing the system~\eqref{goveq2025} to a single ordinary differential equation (ODE) in terms of the travel wave velocity profile \( F(\xi) \). The resulting second-order ODE takes the form:
\begin{equation}
(\tau + \alpha^2) F''(\xi) + \left( (-c^2 + \Gamma_1) F(\xi) + \frac{3c}{2} F(\xi)^2 - \frac{1}{2} F(\xi)^3 \right) = \Gamma_2,
\label{eq:reduced_second_order}
\end{equation}
where \( \Gamma_1 \) and \( \Gamma_2 \) are integration constants emerging from the reduction process. This equation governs the structure of the traveling wave velocity \( F(\xi) \).

\section{Application of WBK Model}\label{sec55}
One of the key applications of the WBK system lies in the detailed study of tsunami dynamics. For tsunami wave modeling using Eq.~\eqref{eq:reduced_second_order}, the specification of boundary conditions is critical for accurately capturing the essential physical processes associated with tsunami generation, propagation, and interaction with coastal regions. Tsunamis are generally initiated by sudden seabed displacements, which induce localized disturbances in the ocean surface. These disturbances evolve into long-wavelength waves that travel across the ocean with small amplitudes in deep water, but grow significantly in height due to nonlinear shoaling effects as they approach the shore.

To describe this behavior, we introduce the traveling wave coordinate transformation \( \xi = x-ct \), where \( c \) is the wave speed. The spatial computational domain is defined over \( \xi \in [0, L] \), where \( \xi = 0 \) represents the offshore wave generation region and \( \xi = L \) denotes the nearshore or coastal boundary. The governing reduced system involves two dependent variables: \( F(\xi) \), denoting the horizontal fluid velocity, and \( G(\xi) \), the free surface elevation.

At the offshore boundary \( \xi = 0 \), the fluid is initially at rest, and the surface elevation is prescribed based on seabed displacement. Therefore, we impose:
\[
F(0) = 0, \quad G(0) = G_{\text{in}},
\]
where \( G_{\text{in}} \) represents the initial free surface displacement. To ensure a smooth and symmetric onset of the wave and to avoid introducing artificial gradients into the numerical model, we also enforce:
\[
F'(0) = 0, \quad G'(0) = 0.
\]
However, in the vicinity of the coast, the wave undergoes steepening, which is reflected in a non-zero velocity gradient. To account for this physically realistic behavior, we prescribe:
\[
F'(L) = F_0,
\]
where \(F_0\neq 0 \) characterizes the sharp transition in velocity near the shore.

Together, these five boundary conditions:
\[
F(0) = 0, \quad G(0) = G_{\text{in}}, \quad F'(0) = 0, \quad G'(0) = 0, \quad F'(L) = F_0,
\]
are carefully selected to model the full evolution of a tsunami wave, from its generation in the deep ocean to its amplification near the shoreline. This boundary framework supports the stable and accurate implementation of numerical techniques such as the shooting method or finite difference schemes on a finite computational domain. By using the above boundary conditions, the value of the integral constant can be determined as follows: $\Gamma_1=G_{\text{in}}$ and $\Gamma_2=0.$ Hence, the final boundary value problem becomes 
\begin{equation}
(\tau + \alpha^2) F''(\xi) + \left( (-c^2 +G_{\text{in}} ) F(\xi) + \frac{3c}{2} F(\xi)^2 - \frac{1}{2} F(\xi)^3 \right) = 0,
\label{eq:reduced_second_order2025}
\end{equation}
subject to the boundary conditions $F(0)=0, F'(L)=F_0.$ The obtained numerical results are discussed in detail graphically as follows:

\begin{figure}[htbp]
    \centering
    \begin{subfigure}[b]{0.48\textwidth}
        \centering
        \includegraphics[width=\textwidth]{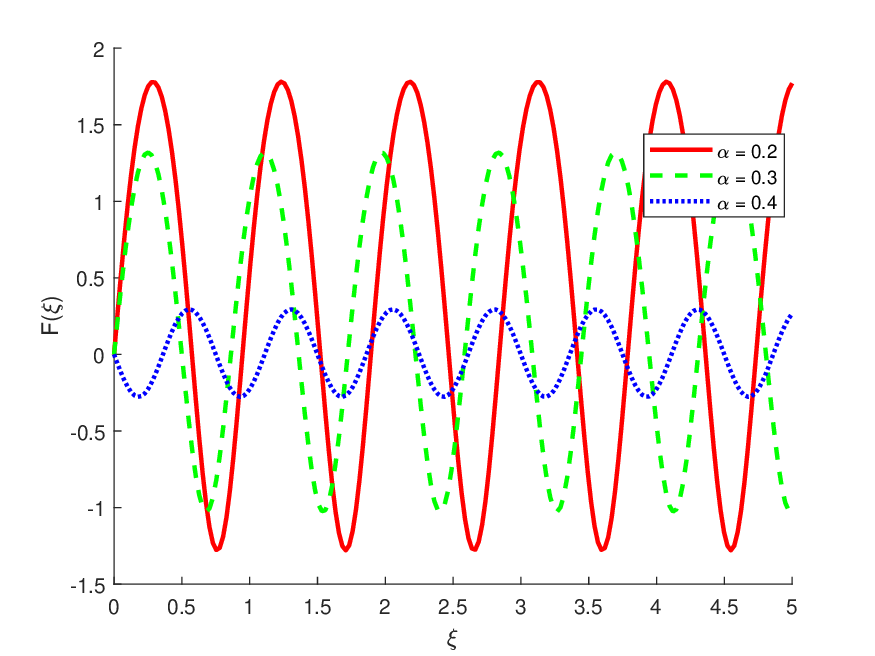}
        \caption{}
        \label{fig:alpha-effect}
    \end{subfigure}
    \hfill
    \begin{subfigure}[b]{0.48\textwidth}
        \centering
        \includegraphics[width=\textwidth]{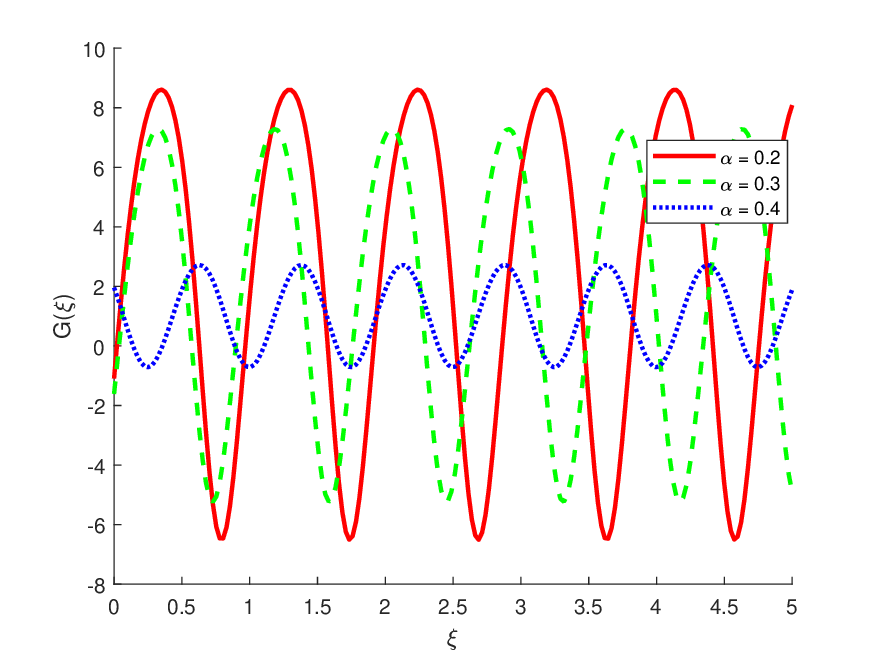}
        \caption{}
        \label{fig:tau-effect}
    \end{subfigure}
    
    \caption{(a) Influence of diffusion parameter \(\alpha\) on travel wave velocity profile  \(u(x,t)\)}, and (b) Influence of diffusion parameter \(\alpha\) on travel wave free surface elevation $v(x,t)$.  Note that the other parameters are fixed $c=5, \tau=-0.5, L=5, \mbox{$G_{in}=1$, $F_0=1$}$.
    \label{fig:fig20251}
\end{figure}
Figure~\ref{fig:fig20251} demonstrates the influence of the diffusion parameter \( \alpha \) on the structure of traveling wave solutions in a one-dimensional shallow water flow modeled by the WBK system. In subfigure~(a), the axial velocity profile \( u(x,t) \approx F(\xi) \) exhibits a distinct dependence on \( \alpha \): as the diffusion parameter increases, the sharpness of the velocity peak near the wave crest decreases, and the entire profile becomes more flattened and broadened. This behavior is physically intuitive, as a higher diffusion parameter represents stronger momentum diffusion or dispersive regularization, which suppresses sharp gradients in the velocity field and promotes smoother flow structures. Similarly, subfigure~(b) shows the corresponding free surface elevation \( v(x,t) \approx G(\xi) \), where increasing \( \alpha \) leads to a more gentle and spread-out wave crest. The flattening of the free surface can be attributed to the influence of \( \alpha \) through both the smoothing of the axial velocity and the added dispersive correction \( -\alpha F'(\xi) \) in the expression for \( G(\xi) \). Physically, this reflects the tendency of dispersive or viscous effects in shallow water systems to distribute wave energy more evenly across space, thereby reducing the nonlinearity and sharpness typically associated with undisturbed solitary wave profiles. Overall, the figure captures the regularizing role of diffusion in modifying both the flow velocity and surface shape of traveling shallow water waves.

\begin{figure}[htbp]
    \centering
    \begin{subfigure}[b]{0.48\textwidth}
        \centering
        \includegraphics[width=\textwidth]{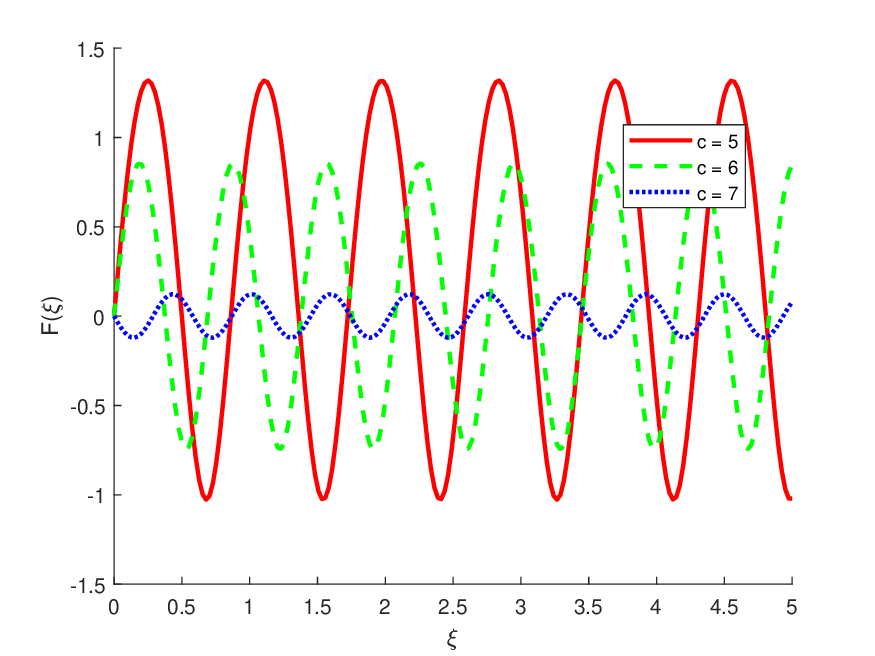}
        \caption{}
        \label{fig:alpha-effect}
    \end{subfigure}
    \hfill
    \begin{subfigure}[b]{0.48\textwidth}
        \centering
        \includegraphics[width=\textwidth]{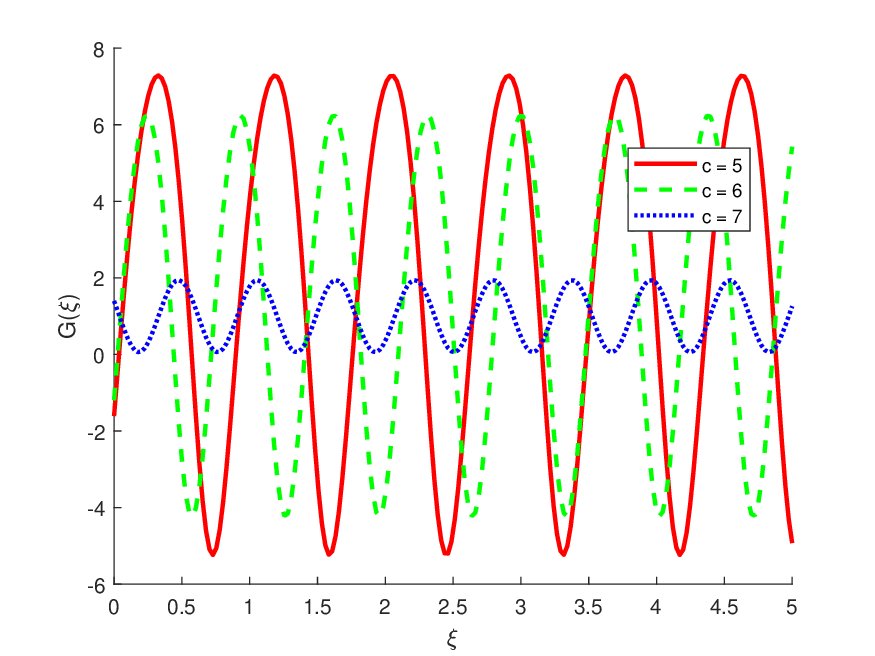}
        \caption{}
        \label{fig:tau-effect}
    \end{subfigure}
    
    \caption{(a) Effect of wave speed \(c\) on travel wave velocity profile  \(u(x,t)\)}, and (b) Effect of the wave speed \(c\) on travel wave free surface elevation $v(x,t)$. Note that the other parameters are fixed $\alpha=0.3, \tau=-0.5, L=5, \mbox{$G_{in}=1$, $F_0=1$}.$
    \label{fig:20252}
\end{figure}

Figure~\ref{fig:20252} illustrates the effect of the wave speed \( c \) on the traveling wave profiles for axial velocity and free surface elevation in a one-dimensional shallow water system governed by the WBK equation. In subfigure~(a), the velocity profile \( u(x,t) \approx F(\xi) \) shows a clear trend with increasing \( c \): as wave speed increases, the peak magnitude of the velocity decreases and the profile broadens. Physically, this behavior is expected because in the wave-fixed frame, a faster-moving wave causes the relative fluid velocity to appear smaller, effectively reducing the amplitude of \( F(\xi) \). Moreover, a higher wave speed enhances convective effects, stretching the wave structure and leading to smoother transitions in the velocity field. In subfigure~(b), the corresponding free surface elevation \( v(x,t) \approx G(\xi) \) also becomes broader and less steep as \( c \) increases. This is due to the fact that \( G(\xi) = cF(\xi) - \frac{1}{2}F(\xi)^2 - \alpha F'(\xi) + \Gamma_1 \), so changes in both the amplitude and gradient of \( F(\xi) \) directly influence the surface profile. The decreasing peak in surface elevation with increasing wave speed reflects the redistribution of wave energy over a wider region, characteristic of dispersive-dominated wave propagation. Overall, these figures highlight the role of wave speed in modulating the shape, steepness, and energy concentration of both velocity and free surface profiles in traveling shallow water waves.

\begin{figure}[htbp]
    \centering
    \begin{subfigure}[b]{0.48\textwidth}
        \centering
        \includegraphics[width=\textwidth]{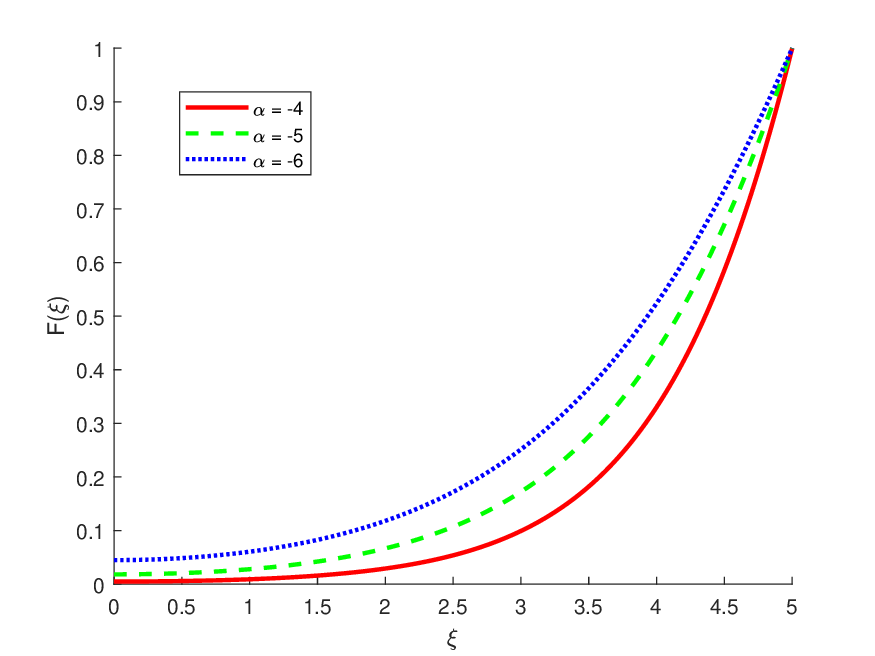}
        \caption{}
        \label{fig:alpha-effect}
    \end{subfigure}
    \hfill
    \begin{subfigure}[b]{0.48\textwidth}
        \centering
        \includegraphics[width=\textwidth]{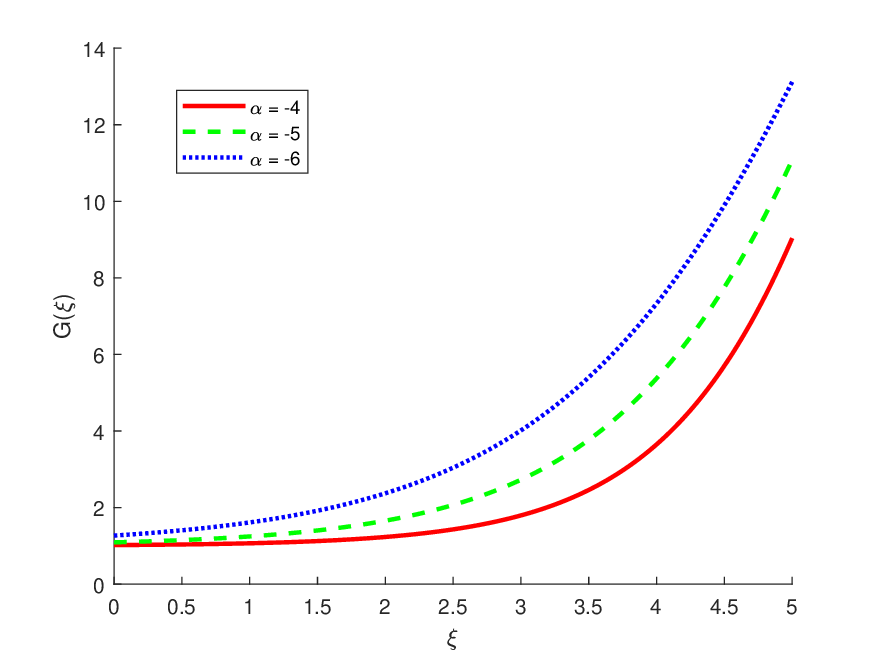}
        \caption{}
        \label{fig:tau-effect}
    \end{subfigure}
    
    \caption{(a) Effect of larger values of $\alpha$ on travel wave velocity profile  \(u(x,t)\)}, and (b) Effect of larger values of $\alpha$ on travel wave free surface elevation $v(x,t)$. Note that the other parameters are fixed $c=5, \tau=-0.5, L=5, \mbox{$G_{in}=1$, $F_0=1$}.$
    \label{fig:20253}
\end{figure}
Figure~\ref{fig:20253} illustrates the influence of the larger negative values of the dispersive parameter $\alpha$ on the traveling wave solution of the WBK equations, focusing on both the travel wave velocity field $F(\xi)$ and the free surface elevation $v(x,t)$. In subplot~6(a), as the dispersion parameter $\alpha$ becomes more negative (i.e., its magnitude increases), the velocity profile $F(\xi)$ becomes sharper and more peaked. This indicates that stronger negative dispersion tends to concentrate the wave energy, resulting in a narrower wave with a higher peak. In contrast to classical dispersion that spreads waves, here the negative dispersion acts in a focusing manner, enhancing the steepness of the velocity crest. From figure~6(b), a similar type of behavior for the travel wave free surface elevation $G(\xi)$ is also observed for different larger negative values of $\alpha$. This behavior highlights the role of dispersive terms in shaping the wave structure in nonlinear shallow water models like the WBK system.

The results obtained from the analysis of diffusion and wave speed effects have significant applications in tsunami wave modeling. The observed smoothing of both axial velocity and free surface elevation with increasing diffusion parameter $\alpha$ highlights the importance of incorporating dispersive effects in tsunami simulations to capture realistic wave spreading and avoid nonphysical steep gradients. Similarly, the influence of wave speed $c$ demonstrates that faster-traveling tsunamis exhibit broader and less steep wave profiles in the traveling frame, which is critical for predicting wave arrival times and energy distribution. These findings support more accurate modeling of tsunami generation, propagation, and coastal interaction by guiding the selection of physically appropriate initial and boundary conditions. Overall, the results enhance our understanding of how tsunami wave characteristics evolve in space and time, contributing to improved early warning systems and inundation risk assessment.


In the next section, we explore conservation laws for WBK differential equations using the direct multiplier method. This method involves finding suitable multipliers that convert the given PDEs into a total divergence form, valid for all its solutions. These multipliers are functions of the independent variables, dependent variables, and their derivatives, which are determined by solving an overdetermined system. Once identified, they yield conserved densities and fluxes associated with the equation. Conservation laws are crucial as they represent invariant physical quantities like mass, momentum, and energy. They offer deep insight into the structure of the system and serve as valuable tools for validating analytical and numerical solutions, ensuring accuracy and stability in simulations.

\section{Conservation laws}\label{sec4}

Let's define the following notations: 
\begin{equation}\label{eq12}
R^1[u,v] \triangleq u_t + u u_x + v_x + \alpha u_{xx} = 0,    
\end{equation}
\begin{equation}\label{eq13}
R^2[u,v] \triangleq  v_t + (u v)_x + \tau u_{xxx} - \alpha v_{xx} = 0.    
\end{equation}

A set of local variables dependent divergence-type conservation laws for the system of partial differential equations \eqref{eq12} and \eqref{eq13} can be derived in the following form:
\[
D_i \psi_i[u,v] \triangleq D_1 \psi_1[u,v] + D_2 \psi_2[u,v] = 0,
\]  
which holds for the solutions manifold of the system of partial differential equations \eqref{eq12} and \eqref{eq13}, where  $\psi_1[u,v], \psi_2[u,v]$ are the flux functions. Further, $D_1$ and $D_2$ stand for the total derivative operators with respect to $ t$ and $x$, respectively. 

A set of multipliers  
$\Lambda^1[u, v] = \Lambda^1(x, t, u, v),~ \Lambda^2[u, v]$ $ = \Lambda^2(x, t, u, v)$ are called the conservation law multipliers if they satisfying  
$$\Lambda^1[u, v] R^1[u, v] +  \Lambda^2[u, v]  R^2[u, v]  \equiv D_i \psi_i[u, v],$$ 
for every solutions \( u, v \) of the system of equations \eqref{eq12} and \eqref{eq13}. Following the methodology systematically outlined in Ref.~\cite{ref30}, we obtain the following results: 
\begin{theorem}\label{thm5}
Based on the divergence expression \( D_i \psi_i[u, v] \), the following equality holds:
\[
E_u(D_i \psi_i[u, v]) \equiv 0, ~\mbox{and}~ E_v(D_i \psi_i[u, v]) \equiv 0,
\]
where \( E_u, E_v \) are the Euler operators defined by  
\[
E_u = \frac{\partial}{\partial u} - D_i \frac{\partial}{\partial u_i} + \cdots 
+ (-1)^s D_{i_1} \cdots D_{i_s} \frac{\partial}{\partial u_{i_1 \cdots i_s}} + \cdots,
\]
\[
E_v = \frac{\partial}{\partial v} - D_i \frac{\partial}{\partial v_i} + \cdots 
+ (-1)^s D_{i_1} \cdots D_{i_s} \frac{\partial}{\partial v_{i_1 \cdots i_s}} + \cdots, ~\mbox{with}~ i, i_1, i_2,....,i_k \in \{x, t\}.
\]
\end{theorem}

\begin{theorem}\label{thm6}
A divergence expression for the system \eqref{eq12}-\eqref{eq13} can be derived by a set of conservation law multipliers \( \Lambda^1(x, t, u, v), \Lambda^2(x, t, u, v) \)  
if and only if  
\begin{equation}\label{eq14}
E_u \big( \Lambda^1[u, v] R^1[u, v] + \Lambda^2[u, v] R^2[u, v] \big) \equiv 0, ~E_v \big( \Lambda^1[u, v] R^1[u, v] + \Lambda^2[u, v] R^2[u, v] \big) \equiv 0
\end{equation}
holds for every solution \( u \) of the system \eqref{eq12}-\eqref{eq13} .
\end{theorem}

Utilizing Theorem \ref{thm6}, we obtain the following theorem that provides the desired local conservation laws associated with equations \eqref{eq3}.

\begin{theorem}\label{thm7}
Under two arbitrary constants \( \alpha, \tau \), the system \eqref{eq3} admits the local conservation laws:
\[
D_t (u) + D_x (\frac{1}{2} u^2 + \alpha u_x + v) = 0, 
\]
\[
D_t (v) + D_x (-\alpha v_x + \tau u_{xx} + u v) = 0, 
\]

\[
D_t (u v) + D_x (u^2 v - u \alpha v_x + u_x v \alpha + \frac{1}{2} v^2+ u \tau u_{xx} - \frac{1}{2} \tau u_x^2) = 0, 
\]
\[
D_t (-t u v + v x) + D_x (-u^2 t v + u t \alpha v_x- u_x t v \alpha - \frac{1}{2} t v^2 + u x v - u t \tau u_{xx} + \frac{1}{2} t \tau u_x^2- \tau u_x + u_{xx} x \tau + v \alpha - v_x x \alpha) = 0, 
\]
under the set of pairs of conservation law multipliers $\{(1, 0), (0, 1), (v, u), ( -t v, -t u + x) \}$, respectively.  
These are the complete list of local conservation laws obtained by the direct multiplier method for Eq. \eqref{eq3}.
\end{theorem}

\textbf{Proof.} From Eq. \eqref{eq14}, we can derive the system \[
\frac{\partial^2 \Lambda^2}{\partial t^2} = 0, \quad 
\frac{\partial^2 \Lambda^1}{\partial v^2} = 0, \quad 
\frac{\partial \Lambda^1}{\partial x} = 0, \quad
\frac{\partial \Lambda^2}{\partial x} = -\frac{\partial \Lambda^2}{\partial t} \frac{1}{u}, \quad 
\frac{\partial \Lambda^1}{\partial t} = \frac{\partial \Lambda^2}{\partial t} \frac{v}{u}, \quad
\frac{\partial \Lambda^1}{\partial u} = 0, \quad 
\frac{\partial \Lambda^2}{\partial u} = \frac{\partial \Lambda^1}{\partial v}, \]
\[
\frac{\partial \Lambda^2}{\partial v} = 0.
\] 
Then, we have the following solution:
\[
\Lambda^1(x, t, u, v) = (-d_1 t + d_3)v + d_4, \quad
\Lambda^2(x, t, u, v) = (-t u + x)d_1 + d_3 u + d_2
\]
where \( d_i \) (\( i = 1, 2, 3, 4 \)) are all arbitrary constants. These multipliers give the set of conservation laws given by Theorem \ref{thm7}. These conservation laws are very useful to derive the nonlocally related potential systems, which are used to find the nonlocal symmetries. These symmetries are very important for solving complex boundary value problems.

\section{Conclusion}\label{sec5}
In this study, we investigated the exact solutions for Whitham-Broer-Kaup (WBK) equations, a significant model for shallow water wave dynamics that incorporates both nonlinear and dispersive effects. Utilizing Lie symmetry analysis, we constructed an optimal system of one-dimensional subalgebras and derived a new class of analytical soliton solutions in hyperbolic, trigonometric, and rational forms, solutions not previously reported using existing studies. Further, this study shows that symmetry analysis can reduce the complexity of the given governing PDE model by converting it into simpler ODEs, which are comparatively easier to solve. We have also explored the numerical solutions using the shooting method for the reduced system and graphically shown the effect of the wave speed $c$ and $\alpha$ on the velocity profile and free surface elevation. This study demonstrates that the WBK model can effectively simulate tsunami wave dynamics, highlighting its potential for practical applications in coastal hazard prediction and early warning systems. Additionally, the direct multiplier method is employed to obtain a comprehensive set of local conservation laws, reinforcing the physical integrity of the model. These findings contribute to a deeper understanding of nonlinear wave behavior in shallow water and establish a solid framework for further analytical exploration of related nonlinear evolution equations.

\section*{Funding Declaration}
SM acknowledges the financial support provided by the University Grants Commission, India, through the CSIR-UGC NET fellowship for pursuing the doctoral degree.

\section*{Author Contributions Statement}

{\bf First author}: Methodology, Investigation, Formal analysis, Writing—original draft, and Software. {\bf Second author}: Conceptualization, Formal analysis, Writing—review and editing.

\section*{Conflict of interest}
The authors declare that they have no conflict of interest.

\section*{Data availability statement}
All the data that support the findings of this investigation are available within the article.

\bibliographystyle{unsrt}  
\bibliography{references}

\end{document}